\documentstyle[12pt]{article}

\begin{document} 

\begin{titlepage}

\hrule 
\leftline{}
\leftline{
          \hfill   \hbox{\bf CHIBA-EP-95}}
\leftline{\hfill   \hbox{\bf OUTP-96-31P}}
\leftline{\hfill   \hbox{hep-th/9608133}}
\leftline{\hfill   \hbox{June 1996}}
\vskip 5pt
\hrule 
\vskip 1.5cm

\centerline{\large\bf 
An Exact Solution of Gauged Thirring Model 
}
\centerline{\large\bf
in Two Dimensions
$^*$
}   
\vskip 1cm

\centerline{{\bf 
Kei-Ichi Kondo$^{\dagger}$
}}  
\vskip 4mm
\begin{description}
\item[]{\it  
  Theoretical Physics, 
  University of Oxford,
  1 Keble Road, Oxford, OX1 3NP, UK.$^\ddagger$
  }
\item[]{$^\dagger$ 
  E-mail: kondo@cuphd.nd.chiba-u.ac.jp;
  kondo@thphys.ox.ac.uk}    
\end{description}
\vskip 1cm

\centerline{{\bf Abstract}} \vskip .5cm

In two space-time dimensions, we write down the exact and
closed Schwinger-Dyson equation for the gauged Thirring
model which has been proposed recently by the author.
The gauged Thirring model is a natural gauge-invariant
extension of the Thirring model and reduces to the
Schwinger model (in the Abelian case) in the strong
four-fermion coupling limit. The exact SD equation is
derived by making use of the transverse Ward-Takahashi
identity as well as the usual (longitudinal) Ward-Takahashi
identity.  Moreover the exact solution of the SD equation
for the fermion propagator is obtained together with the
vertex function in the Abelian gauged case. 
Finally we discuss the dynamical fermion mass generation
based on the solution of the SD equation.



\vskip 0.5cm
\hrule  


\begin{description}
\item[]{
$^\ddagger$
Address from March 1996 to December 1996.
  On leave of absence from: 
  Department of Physics, Faculty of Science,
  Chiba University, Chiba 263, Japan.
  }
\end{description}

\end{titlepage}

\pagenumbering{roman}
\pagenumbering{arabic}

\newpage

\section{Introduction} 
\setcounter{equation}{0}
It is well known that the existence of some symmetry
in quantum field theory leads to the various identities
which hold among the Green functions and the vertex
functions, which are generally called the Ward-Takahashi
(WT) identities \cite{WT}. 
In the previous paper \cite{Kondo96b}, we have rederived,
based on the path integral formalism, a new type of WT
identities, so-called the transverse WT identities
\cite{Takahashi86}, for the vector current and the axial
vector current.  We have suggested to use the
transverse WT identity as well as the usual (longitudinal)
WT identity to specify the vertex function in the
Schwinger-Dyson (SD) equation in the gauge theory.
Especially, we have shown that they lead to the exact
and closed SD equation for the fermion propagator in Abelian
gauge theory defined on the 1+1 dimensional Minkowski
space-time, when the bare fermion mass $m_0$ is zero
(chiral limit).  Actually, the formal solution of
the SD equation obtained in such a way in massless QED$_2$
agrees with the exact solution obtained from the
path integral formalism. Therefore, the framework of the
SD equation supplemented with the transverse as well as the
longitudinal WT identities can give the exact solution at
least for the Abelian gauge theory in 1+1 dimensional
Minkowski space-time.
\par
In this paper we apply this strategy to solve the Thirring
model \cite{Thirring} and the gauged Thirring model
\cite{Kondo96a} in 1+1 dimensions. First of all, we must
rewrite the Thirring model as a gauge theory.  Such a
procedure has been recently presented, first from the
viewpoint of the hidden local symmetry
\cite{IKSY95} and then from the constrained system of
Batalin-Fradkin type, see 
\cite{Kondo95th1,Kondo95th2,IKN96}.
In this reformulation, the original Thirring model can be
regarded as a gauge-fixed version (unitary gauge) of the
reformulated Thirring model as a gauge theory.
The auxiliary vector field (which is originally introduced
into the theory in order to bi-linearize the four-fermion
interaction) can be identified with the gauge-boson field
in this reformulation, although the corresponding kinetic
term is absent in this stage.  On the other hand, it has
been shown that the kinetic term for the gauge boson field
can be generated due to radiative corrections
\cite{Hands95,IKSY95}.
Quite recently, the Thirring model as a gauge theory has
been further extended so as to include the kinetic term for
the gauge field from the very beginning, which we call the
gauged Thirring model \cite{Kondo96a}. In the gauged
Thirring model, the Thirring model is recovered in the
strong gauge-coupling limit 
$e^2 \rightarrow \infty$.  
On the other hand, the strong four-fermion coupling limit 
$G_T \rightarrow \infty$ reduces to the QED  or QCD 
when the gauge theory in question is Abelian type or
non-Abelian type, respectively.
\par
Then, applying the reformulation of the Thirring
model as a gauge theory to the 1+1 dimensional case, we can
write down the exact SD equation for the fermion
propagator, according to the procedure proposed in the
previous paper
\cite{Kondo96b}.  
In this framework the exact formal solution is obtained
(only) in the chiral limit $m_0=0$. This formal solution
has no extra pole at $p^2 \not= 0$ which implies that the
fermion remains massless for $m_0=0$, which is the same
situation as the massless Schwinger model (QED$_2$)
\cite{Schwinger62}. As we mentioned in the above,
QED$_2$ is regarded as the strong four-fermion coupling
limit 
$G_T \rightarrow \infty$ 
of the Abelian-gauged Thirring model, in which the dynamical
fermion mass generation is expected to occur. Therefore, we
expect that the SD equation should have a solution
corresponding to the dynamical fermion mass generation even
in the limit $m_0=0$. From this viewpoint, we study  the
dynamical fermion mass generation in the pure Thirring
model limit
$e^2 \rightarrow \infty$
and give perspectives of future studies for the gauged
Thirring model with finite $e^2$ ($0 < e^2 < \infty$).

\section{Thirring model as a gauge theory} 
\setcounter{equation}{0}

We consider the Thirring model with the Lagrangian
\begin{eqnarray}
 {\cal L}_{Th} 
 &=& \bar \Psi^j i \gamma^\mu \partial_\mu \Psi^j 
 - \bar \Psi^j \hat{m}_0 \Psi^j
- {G_T \over 2}
(\bar \Psi^j  \gamma_\mu \Psi^j)
(\bar \Psi^k  \gamma^\mu \Psi^k),
\label{ATh}
\end{eqnarray}
where $\Psi^j$ is a Dirac fermion with a flavor index $j$
running from $1$ to $N_f$ and $\hat{m}_0$ denotes the mass
matrix for the fermion. By introducing the auxiliary vector
field $A_\mu$, the Thirring model is rewritten as
\begin{eqnarray}
 {\cal L}_{Th'}
 &=& \bar \Psi^j i \gamma^\mu  
 (\partial_\mu - i A_\mu) \Psi^j 
 -  \bar \Psi^j \hat{m}_0 \Psi^j
 + {1 \over 2G_T} A_\mu^2 .
\label{ATh'}
\end{eqnarray}
By introducing a scalar field $\theta$,
we can write down a gauge-invariant generalization of the
Thirring model as \cite{Kondo96a}
\begin{eqnarray}
 {\cal L}_{Th''} 
 =  \bar \Psi^j i \gamma^\mu  
 (\partial_\mu - i A_\mu) \Psi^j 
 -  \bar \Psi^j \hat{m}_0 \Psi^j
 + {1 \over 2G_T} (A_\mu - \partial_\mu \theta)^2
 - {1 \over 4e^2} F^{\mu\nu}F_{\mu\nu} ,
\label{ATh''}
\end{eqnarray}
where we have also included the kinetic term for the
vector field $A_\mu$ with a gauge coupling constant $e$.
Indeed, the Lagrangian (\ref{ATh''}) is invariant under
the local $U(1)$ gauge transformation:
\begin{eqnarray}
 \Psi(x) &\rightarrow& \Psi(x) e^{i \omega(x)} ,
 \nonumber\\
A_\mu(x) &\rightarrow& A_\mu(x) 
 +  \partial_\mu \omega(x) ,
 \nonumber\\
\theta(x) &\rightarrow& \theta(x) + \omega(x), 
\quad
( \phi(x) \rightarrow \phi(x) e^{i \omega(x)}  )
 \label{gt}
\end{eqnarray}
for a new variable
$
 \phi(x) = e^{i \theta(x)}.
$
\footnote{
The gauged Thirring model can also be regarded with the
Higgs-Kibble model (or gauged non-linear sigma model) in
the presence of fermions: 
\begin{eqnarray}
 {\cal L}_{Th'''}   
 =  \bar \Psi^j i \gamma^\mu  
 (\partial_\mu - i A_\mu) \Psi^j 
 -  \bar \Psi^j \hat{m}_0 \Psi^j
 + {1 \over 2G_T} |(\partial_\mu + i A_\mu) \phi |^2
 - {1 \over 4e^2} F^{\mu\nu}F_{\mu\nu} .
 \label{gNLs}
\end{eqnarray}
}

\par
The original Thirring model is identified with the strong
gauge-coupling limit $e^2=\infty$ of a gauge-fixed version
of the gauge theory with the Lagrangian (\ref{ATh''}),
which we call the {\it (Abelian-)gauged Thirring model}
\cite{Kondo96a}.  
Indeed, the Lagrangian (\ref{ATh''}) reduces to the
Lagrangian (\ref{ATh'}), if we take the unitary gauge:
$\theta(x) \equiv 0$ and set $e^2=\infty$.
However, actual calculations such as loop calculations are
generally impossible in the unitary gauge.  For such
purposes the covariant gauge is most convenient, although
both theories should give the same results on the
gauge-invariant quantities, e.g., the chiral condensate
$\langle \bar \Psi \Psi \rangle$.
\par
Now we briefly review the covariantly
gauge-fixed BRST formulation of the gauged Thirring model,
see \cite{Kondo96a} for more details. It is well known that
the BRST invariant total Lagrangian, 
\begin{eqnarray}
 {\cal L}_{Th'''} = {\cal L}_{Th''} + {\cal L}_{GF} 
+ {\cal L}_{FP},
\end{eqnarray}
is obtained by adding the gauge-fixing term and the
Faddeev-Popov (FP) ghost term
${\cal L}_{GF+FP}$ to the Lagrangian ${\cal L}_{Th''}$.
The nilpotent BRST transformation ($\delta_B^2 * = 0$) is
given by
\begin{eqnarray}
 \delta_B \Psi^j(x) &=&  i C(x) \Psi^j(x),
\nonumber\\ 
 \delta_B A_\mu(x) &=& \partial_\mu C(x),
\nonumber\\ 
 \delta_B \theta(x) &=&   C(x) ,
\nonumber\\ 
 \delta_B B(x) &=& 0,
\nonumber\\ 
 \delta_B C(x) &=& 0 ,
\nonumber\\ 
 \delta_B \bar C(x) &=& i  B(x) .
 \label{BRST}
\end{eqnarray}
Then the BRST invariance of the additional term
is guaranteed by the construction:
\begin{eqnarray}
 {\cal L}_{GF+FP}[A, \theta, C, \bar C,B]
&=& - i \delta_B(\bar C (F[A,\theta]+{\xi \over 2}B)) 
\nonumber\\
&=&   B F[A,\theta]+{\xi \over 2} B^2
 + i \bar C \delta_B F[A,\theta] ,
\end{eqnarray}
where $F[A,\theta]$ is the gauge-fixing condition.
After integrating out the $B$ field, 
we obtain
\begin{eqnarray}
 {\cal L}_{GF}
 &=& - {1 \over 2\xi} (F[A,\theta])^2,
\\
 {\cal L}_{FP} &=&   i \bar C \delta_B F[A,\theta] 
 =   i \bar C \left( {\delta F[A,\theta]
 \over \delta A_\mu} \partial_\mu C(x) 
 + {\delta F[A,\theta] \over \delta \theta}
 C  \right).
\end{eqnarray}
The covariant gauge is given by
\begin{eqnarray}
 F[A,\theta] = \partial^\mu A_\mu ,
\end{eqnarray}
leading to
\begin{eqnarray}
{\cal L}_{FP}
 =  i \bar C  \partial^\mu \partial_\mu  .
\end{eqnarray}
\par
For our purposes, the $R_\xi$ gauge is more convenient:
\begin{eqnarray}
 F[A,\theta] = \partial^\mu A_\mu + \xi G^{-1} \theta,
 \label{gauge}
\end{eqnarray}
so that the  crossing term  
$
- G^{-1} A_\mu \partial_\mu \theta
$ 
coming from
$
 {1 \over 2G_T} (A_\mu - \partial_\mu \theta)^2
$ is
canceled with that of the gauge-fixing term 
${\cal L}_{GF}$ and the total Lagrangian 
${\cal L}_{Th'''}$ is decomposed
into three parts:
\begin{eqnarray}
  {\cal L}_{Th'''} 
 &=& {\cal L}_{\Psi,A} + {\cal L}_{\theta}+ {\cal L}_{FP} ,
 \nonumber\\
{\cal L}_{\Psi,A}
&=& \bar \Psi^j i \gamma^\mu 
(\partial_\mu - ie A_\mu) \Psi^j 
 -  \bar \Psi^j \hat{m}_0 \Psi^j
\nonumber\\&&
 + {e^2 G_T^{-1} \over 2}(A_\mu)^2
 - {1 \over 2\xi}(\partial^\mu A_\mu)^2
 - {1 \over 4}  F_{\mu\nu}F^{\mu\nu} ,
\nonumber\\
{\cal L}_{\theta}
&=&  {G_T^{-1} \over 2}(\partial_\mu \theta)^2
 - {\xi \over 2} (G_T^{-1} \theta)^2,
\nonumber\\
 {\cal L}_{FP} 
&=&  i \bar C \left( \partial_\mu \partial^\mu
 + \xi G_T^{-1} \right) C ,
 \label{L}
\end{eqnarray} 
where we have rescaled the field $A_\mu$ as
$A_\mu \rightarrow eA_\mu$.
Note that the scalar field $\theta$ is completely decoupled
independently of $\xi$ in the $R_\xi$ gauge.  This is an
advantage of taking the $R_\xi$ gauge in this paper.
\par
We find the following limiting cases \cite{Kondo96a}.
In the limit $e^2 \rightarrow \infty$, this model reduces to
the gauge invariant-reformulation of the (massive) Thirring
model \cite{IKSY95} where $\theta$ is nothing but the
well-known St\"uckelberg field (or the Batalin-Fradkin
field). In the limit $G_T \rightarrow \infty$, QED with
$N_f$ flavors is recovered:
$
 {\cal L}_{QED}  
 =  \bar \Psi^j i \gamma^\mu 
(\partial_\mu - i e A_\mu) \Psi^j 
 -  \bar \Psi^j \hat{m}_0 \Psi^j
 - {1 \over 4} F^{\mu\nu}F_{\mu\nu} .
$
The weak gauge-coupling limit $e \rightarrow 0$ of the
gauged Thirring model reduces to the non-linear $\sigma$
model.
$
 {\cal L}_{NL\sigma}  =  \kappa |\partial_\mu  \phi |^2
\quad (|\phi | = 1) .
$

\section{WT identity and SD equation} 
\setcounter{equation}{0}

The usual (longitudinal) WT identity
for the vector current
\begin{eqnarray}
{\cal J}_\mu(x) := \bar \Psi(x) \gamma_\mu \Psi(x)  
\end{eqnarray}
is given by
\begin{eqnarray}
\partial_\mu \langle {\cal J}^\mu(x) ;
 \Psi(y) ; \bar \Psi(z) \rangle_c
&=&  \langle {1 \over e} 
\partial_\mu \Delta^{\mu\rho} (\partial) A_\rho(x) ;
 \Psi(y) ; \bar \Psi(z) \rangle_c
  \nonumber\\
 &=&  \langle \Psi(y) \bar \Psi(z) \rangle_c \delta^D(x-z)
 -  \langle \Psi(y) \bar \Psi(z) \rangle_c \delta^D(x-y),
 \label{WTL0}
\end{eqnarray}
where $\Delta^{\mu\rho}$ is the inverse gauge-boson
propagator obtained from (\ref{L}) and
$\langle ...\rangle_c$ denotes the connected
correlation function.
For the proper fermion-boson vertex function in momentum
representation
\begin{eqnarray}
 S(q)\Gamma^\mu(q,p)S(p) := \int d^Dy \int d^Dz 
 e^{i(q \cdot y-p \cdot z)} 
 \langle {\cal J}^\mu(0) ;
 \Psi(y) ; \bar \Psi(z) \rangle_c,
 \label{FT1}
\end{eqnarray}
the well-known form of the WT identity is obtained from
(\ref{WTL0}):
\begin{eqnarray}
  k_\mu \Gamma^\mu(q,p) = S^{-1}(q) - S^{-1}(p) ,
  \quad
  k_\mu := q_\mu - p_\mu ,
 \label{WTL1}
\end{eqnarray}
where $S(p)$ is the full fermion propagator in momentum
representation.
\par
In the previous paper \cite{Kondo96b}, the identity for the
rotation of the vector current, which we call the
{\it transverse} WT identity, has been rederived based on
the path integral formalism. Such a type of WT identities
was first derived by Takahashi
\cite{Takahashi86}.   We have found that, in $D=1+1$
dimensions, the transverse WT identity has the 
remarkably simple form:
\footnote{
In two dimensions, the existence of chiral anomaly does not
change the transverse WT identity \cite{Kondo96b}.
}
\begin{eqnarray}
&& \partial_\mu \langle {\cal J}_\nu(x) ;
\Psi(y) \bar \Psi(z) \rangle_c 
- \partial_\nu 
 \langle {\cal J}_\mu(x)  ;
\Psi(y) \bar \Psi(z) \rangle_c
 \nonumber\\ 
&=&  
\langle \bar \Psi(x) \{ \sigma_{\mu\nu}, \hat{m}_0 \}
\Psi(x);
\Psi(y) \bar \Psi(z) \rangle_c
 \nonumber\\&&
 - \langle \Psi(y) \bar \Psi(x) 
\rangle_c \sigma_{\mu\nu} \delta^D(x-z)
 - \sigma_{\mu\nu} \langle \Psi(x) \bar
\Psi(z)  \rangle_c  \delta^D(x-y) ,
 \label{WTT0}
\end{eqnarray}
where 
\footnote{
In 1+1 dimensional space-time, we choose
$
\gamma^0 = \sigma_2, \gamma^1 =  i \sigma_1, 
\gamma^5 := \gamma^0 \gamma^1 =  \sigma_3,
(\epsilon_{01} = 1),
$
which implies 
$
 \sigma_{\mu\nu} = i \epsilon_{\mu\nu} \gamma_5
$
and
$
\gamma_\mu \gamma_5 = \epsilon_{\mu\nu}\gamma^\nu .
$
}
$
 \sigma_{\mu\nu} := {i \over 2}[ \gamma_\mu, \gamma_\nu ] .
$
In the chiral limit $M=0$, especially, the transverse
WT identity leads to the surprisingly simple identity
for the rotation of the vector vertex in $D=2$ dimensions:
\begin{eqnarray}
  k_\mu \Gamma_\nu(q,p) - k_\nu \Gamma_\mu(q,p)
  = S^{-1}(q)\sigma_{\mu\nu} + \sigma_{\mu\nu}S^{-1}(p) ,
 \label{WTT1}
\end{eqnarray}
which is similar in form to the longitudinal part
(\ref{WTL1}).
\par
The bare gauge-boson propagator 
$D^{(0)}_{\mu\nu}(k)$ is written as
\begin{eqnarray}
 D^{(0)}_{\mu\nu}{}^{-1}(k) 
 = (k^2 - e^2 G_T^{-1}) g_{\mu\nu} -  k_\mu k_\nu 
 +  \xi(k^2)^{-1} k_\mu k_\nu ,
\end{eqnarray}
if we adopt the (nonlocal) $R_\xi$ gauge with
(momentum-dependent) gauge-fixing parameter $\xi(k^2)$
\cite{IKSY95,Kondo96a}. 
In momentum representation, the SD equation for the full
gauge-boson propagator $D_{\mu\nu}(k)$ is given by
\begin{eqnarray}
 D_{\mu\nu}^{-1}(k) 
 &=& D^{(0)}_{\mu\nu}{}^{-1}(k) - \Pi_{\mu\nu}(k),
 \nonumber\\
 \Pi_{\mu\nu}(k) &:=& e^2 \int {d^Dp \over (2\pi)^D}
 {\rm tr}[\gamma_\mu S(p) \Gamma_\nu(p,p-k) S(p-k)].
\end{eqnarray}
In the gauge theory, the vacuum polarization tensor
should have  the transverse form:
\begin{eqnarray}
 \Pi_{\mu\nu}(k)  
 =  \left( g_{\mu\nu} - {k_\mu k_\nu \over k^2} \right) 
 \Pi(k),
\end{eqnarray}
as long as the gauge invariance is preserved.
Hence we can write the full gauge-boson propagator in the
following form:
\begin{eqnarray}
 D_{\mu\nu}(k) 
 &=& D_T(k^2) \left( g_{\mu\nu} - {k_\mu k_\nu \over k^2}
\right)
 + {\xi(k^2) \over k^2-\xi(k^2) e^2 G_T^{-1}} 
 {k_\mu k_\nu \over k^2},
 \nonumber\\
 D_T(k^2) &:=& {1 \over k^2 - e^2 G_T^{-1} - \Pi(k)} .
\end{eqnarray}
This decomposition of the full gauge-boson propagator 
into the transverse $D_{\mu\nu}^T(k)$ and the longitudinal
part $D_{\mu\nu}^L(k)$ is most convenient for our purposes,
since this decomposition is preserved in the SD equation
for the fermion propagator as can be seen shortly.
Since there are $N_f$ identical fermions, the vacuum
polarization in 1+1 dimensions is given by
\begin{eqnarray}
  \Pi(k) = N_f {e^2 \over \pi},
\end{eqnarray}
which is consistent with the chiral anomaly in 1+1
dimensions, see \cite{Kondo96b}.  
This implies that the mass $\mu_A$ for the
gauge field  $A_\mu$ is dynamically generated and given by 
$\mu_A = e/\sqrt{\pi}$ .

\par
On the other hand, the SD equation for the full fermion
propagator $S(p)$ is given by
\begin{eqnarray}
 S_0^{-1}(p)S(p) = 1 + ie^2  \int {d^Dk \over (2\pi)^D}
   \gamma^\mu D_{\mu\nu}(k) S(p-k) \Gamma_\nu(p-k,p) S(p),
   \label{SDf}
\end{eqnarray}
where $S_0$ the bare fermion propagator:
\begin{eqnarray}
 S_0(p) := {1 \over \hat{p}- m_0},
 \quad
 (\hat{p} := \gamma^\mu p_\mu) .
\end{eqnarray}

\par
As shown in the previous paper \cite{Kondo96b}, the
transverse and the longitudinal WT identities can specify
the vertex function which appears in the {\it integrand} of
the SD equation for the fermion propagator.
Note that
\begin{eqnarray}
&& D_{\mu\nu}(k) \tilde \Gamma^\nu(q,p)
 \nonumber\\
 &=& D_{\mu\nu}^L(k) \tilde \Gamma^\nu(q,p) 
 + D_{\mu\nu}^T(k) \tilde \Gamma^\nu(q,p)  
 \nonumber\\
 &=& {\xi(k^2) \over k^2-\xi(k^2) e^2 G_T^{-1}}{k_\mu \over
k^2}
 [k_\nu \tilde \Gamma^\nu(q,p)]
 +  D_T(k^2) {k^\nu \over k^2} 
[k_\nu \tilde \Gamma_\mu(q,p) 
- k_\mu \tilde \Gamma_\nu(q,p)],
\label{integrand}
\end{eqnarray}
where we have defined $q := p-k$ and
\begin{eqnarray}
 \tilde \Gamma_\nu(q,p) := S(q) \Gamma_\nu(q,p) S(p) .
\end{eqnarray}
Substituting the longitudinal WT identity (\ref{WTL1}) and
the transverse WT identity (\ref{WTT1}) into
(\ref{integrand}), we get the exact SD equation
\cite{Kondo96b} for the fermion propagator (in the chiral
limit $m_0=0$):
\begin{eqnarray}
  \hat{p} S(p) = 1 + ie^2 \int {d^2k \over (2\pi)^2}
  S(p-k) \hat{k}
  \left[ {D_T(k^2) - L_\xi(k^2) \over k^2} \right]  ,
\label{SDm}
\end{eqnarray}
where $\hat{p} := \gamma^\mu p_\mu$,
$\hat{k} := \gamma^\mu k_\mu$ and 
\begin{eqnarray}
  D_T(k^2) := {1 \over k^2 - e^2 G_T^{-1} - e^2 N_f/\pi},
\quad
  L_\xi(k^2) 
  := {\xi(k^2) \over k^2- \xi(k^2) e^2 G_T^{-1}} .
\label{SDm'}
\end{eqnarray}
This SD equation is exact in 1+1 space-time dimensions.
The solution is easily found by moving to the coordinate
space, since the Fourier transformation changes the
convolution in momentum space into the simple product in
the coordinate space:
\begin{eqnarray}
  i \hat{\partial} S(x) = \delta^2(x) 
  - ie^2 S(x) \int {d^2k \over (2\pi)^2}  \hat{k} 
  \left[ {D_T(k^2) - L_\xi(k^2) \over k^2}  
\right] 
  e^{-i k \cdot x} .
\label{SDc}
\end{eqnarray}
This equation can be solved exactly.
Thus the full fermion propagator of the 1+1 dimensional
(Abelian-)gauged Thirring model in the $R_\xi$ gauge
(\ref{gauge}) is given by
\begin{eqnarray}
  S(x) = S_0(x)
  \exp \left\{ - i e^2 \int {d^2k \over (2\pi)^2}
  \left[ {D_T(k^2) - L_\xi(k^2) \over k^2} 
\right]
  (e^{-i k \cdot x}-1) \right\} ,
  \label{sol}
\end{eqnarray}
where we have assumed the translational invariance
$S(x,y) = S(x-y,0) := S(x-y)$  
and the massless bare fermion propagator $S_0$ is given by
\begin{eqnarray}
S_0(p) := {1 \over \hat{p}}, 
\quad 
S_0(x) = \int {d^2p \over (2\pi)^2} e^{ip \cdot x}S_0(p)
=  {1 \over 2\pi} {\hat{x} \over x^2} .
\end{eqnarray}
\par
In the limit $G_T \rightarrow \infty$, (\ref{sol})
reproduces the exact solution of massless QED$_2$
(Schwinger model) \cite{Schwinger62} (see also the
Appendix of \cite{FM94}):
\begin{eqnarray}
  S(x) = S_0(x)
  \exp \left\{ - i e^2 \int {d^2k \over (2\pi)^2}
  \left[ {1 \over k^2(k^2-N_fe^2/\pi)}   
  - {\xi(k^2) \over k^4} 
\right]
  (e^{-i k \cdot x}-1) \right\} ,
  \label{solqed}
\end{eqnarray}
which agrees also with Stam's result \cite{Stam83} based on
the gauge technique.

\section{Thirring model as strong coupling limit}
\setcounter{equation}{0}
\par
The pure Thirring model is recovered in the strong
gauge-coupling limit 
$e^2 \rightarrow \infty$.  
In this limit, the gauge-independent part $D_T$ reads
\begin{eqnarray}
 e^2 D_T(k^2) \rightarrow
 - {1 \over  G_T^{-1} + N_f/\pi}  
 = - {G_T \over 1 + N_f G_T/\pi} := - \lambda_0.
 \label{renG}
\end{eqnarray}
On the other hand, for the gauge-dependent part $L_\xi$, we
consider the limit 
$\xi \rightarrow \pm \infty$ 
which corresponds to the unitary gauge, $\theta = 0$. This
gauge should reproduce the original Thirring model (at
least in the quenched limit
$N_f \rightarrow 0$ or in the absence of quantum
correction of the gauge-boson propagator).
In the limit $\xi \rightarrow \pm \infty$, we see
$
  L_\xi(k^2) \rightarrow - G_T ,
$
and hence
\begin{eqnarray}
 { e^2 D_T(k^2) - L_\xi(k^2) \over k^2} 
  \rightarrow - {\lambda_\infty \over k^2} ,
\end{eqnarray}
where
\begin{eqnarray}
\lambda_\infty := \lambda_0 - G_T 
= - {N_f G_T^2/\pi \over 1 + N_f G_T/\pi} .
\end{eqnarray}
Thus we get the exact solution in the gauge 
$\xi \rightarrow \pm \infty$ as
\begin{eqnarray}
  S(x) = S_0(x)
  \exp \left\{ i \lambda_\infty\
  \int {d^2k \over (2\pi)^2} {1 \over k^2} 
  (e^{-i k \cdot x}-1) \right\} ,
  \label{solth}
\end{eqnarray}
which agrees with Johnson's result \cite{Johnson61}.
The solution in the $\xi=0$ gauge is obtained by replacing
$\lambda_\infty$ with $\lambda_0$ in (\ref{solth}).

\section{SD equation in the momentum space}
\setcounter{equation}{0}
\par
As proposed in the previous paper \cite{Kondo96b}, 
rather than the inverse propagator
\begin{eqnarray}
    S(p)^{-1} = A(p)\hat{p} - B(p) ,
\end{eqnarray}
it is more convenient to use the following decomposition in
the SD equation for $S(p)$:
\begin{eqnarray}
   S(p) = {\hat{p} X(p) + Y(p) \over p^2},
\end{eqnarray}
and solve a pair of integral equations for $X$ and $Y$.
In general, $X$ and $Y$ couple each other. 
Nevertheless, the equations are still linear in each
variable, $X$ or $Y$, and hence we do not need to take the
linearization approximation, which is in sharp contrast
with the pair of equations for $A$ and $B$ obtained from
the decomposition of the SD equation for $S^{-1}(p)$,
see \cite{Kondo96b}.   The solution, $X$ and $Y$ for such a
pair of equations may include the non-linear effect, since
$X$ and
$Y$ are related with $A$ and $B$ as follows:
\begin{eqnarray}
   X(p) := {A(p)p^2 \over A^2(p)p^2+B^2(p)}, 
   \quad
   Y(p) := {B(p)p^2 \over A^2(p)p^2+B^2(p)}.
\end{eqnarray}
The wavefunction renormalization $A(p)$ and the mass
function $M(p)$ is obtained from $X$ and $Y$ as
\begin{eqnarray}
   M(p) := {B(p) \over A(p)} = {Y(p) \over X(p)}, 
   \quad
   Z(p) := {1 \over A(p)}  
   =  X(p) \left( 1+{M^2(p) \over p^2} \right) .
\end{eqnarray}
The chiral order parameter is determined simply as
\begin{eqnarray}
 \langle \bar \Psi \Psi \rangle/N_f
 = \int {d^2p \over (2\pi)^2} {\rm tr}[S(p)]
 = {\rm tr}(1) \int_0^{\Lambda^2} {dp^2 \over 4\pi} 
  {Y(p) \over p^2} .
\label{op}
\end{eqnarray}

\par
In 1+1 dimensions, especially, the SD equation (\ref{SDm})
is linear in
$S$ and can be reduced to a decoupled pair of integral
equations for $X$ and $Y$:  
\begin{eqnarray}
  X(p) &=&  1 +  \int {d^2k \over (2\pi)^2} 
  {k \cdot (p-k) \over (p-k)^2}
 \left[ {e^2 D_T(k^2) - L_\xi(k^2) \over k^2} \right]
X(p-k),
  \nonumber\\
  Y(p) &=&  \int {d^2k \over (2\pi)^2} 
  {p \cdot k \over (p-k)^2}
 \left[ {e^2 D_T(k^2) - L_\xi(k^2) \over k^2} \right]
Y(p-k),
\label{SD2}
\end{eqnarray}
which can be solved independently.

\section{Dynamical mass generation?}
\setcounter{equation}{0}

To look for the solution representing the dynamical mass
generation, we move back to the Euclidean space and solve
the SD equation (\ref{SD2}) in momentum space.
The solution of (\ref{SD2}) is easily obtained
especially in the gauge
$\xi=0, \pm \infty$.   
After change of variable $p-k=q$, Eq.~(\ref{SD2}) in
the pure Thirring model limit $e^2 \rightarrow \infty$
reads
\begin{eqnarray}
  X(p) &=&  1 +  \int {d^2q \over (2\pi)^2} 
  {(p-q) \cdot q \over q^2}
  \left[ {\lambda  \over (p-q)^2} \right] X(q) ,
  \nonumber\\
  Y(p) &=&   \int {d^2q \over (2\pi)^2}  
  {p \cdot (p-q) \over q^2}
   \left[ {\lambda \over (p-q)^2} \right] Y(q) ,
\label{SD3}
\end{eqnarray}
where $\lambda$ denotes $\lambda_0$ or $\lambda_\infty$
corresponding to the gauge  $\xi=0$ or $\pm \infty$,
respectively. 
After angular integration,
(\ref{SD3}) reduces to
\begin{eqnarray}
  X(p) =  1 -  {\lambda \over 4\pi}  
  \int_{p^2}^\infty {dq^2 \over q^2} X(q) ,
\quad
  Y(p) =   {\lambda \over 4\pi}
  \int_0^{p^2} {dq^2 \over q^2} Y(q) .
\label{SD4}
\end{eqnarray}
These are integral equation of the Volterra
type and can be converted to the first order differential
equation 
\begin{eqnarray}
 {d \over dp^2} X(p) = {\lambda \over 4\pi}{X(p) \over p^2},
 \quad
 {d \over dp^2} Y(p) = {\lambda \over 4\pi}{Y(p) \over p^2},
\end{eqnarray}
with a boundary (or initial) condition, 
$X(\infty)=1, Y(0)=0$. 
The solution of the differential equation is obtained as
\begin{eqnarray}
  X(p) =  C_1 (p^2)^{\lambda \over 4\pi},
  \quad
  Y(p) =  C_2 (p^2)^{\lambda \over 4\pi},  
\end{eqnarray}
with constants $C_1, C_2$, which imply
\begin{eqnarray}
   M(p) := {B(p) \over A(p)} 
   = {C_2 \over C_1} =: M, 
   \quad
   Z(p) := {1 \over A(p)}  
   = X(p)   \left( 1+ {M \over p^2} \right) .
\end{eqnarray}

The infrared (IR) boundary condition for $Y(p)$: $Y(p=0)=0$
is satisfied by an arbitrary constant $C_2=MC_1$. Note that
the dynamical fermion mass function $M(p)$ is
$p$-independent constant $M$.  If there is no cutoff,
$X(p)$ has no finite solution.  By introducing the
ultraviolet (UV) cutoff
$\Lambda$, the UV boundary condition
$X(\Lambda) = 1$ determines the solution uniquely
$C_1=(\Lambda^2)^{-\lambda \over 4\pi}$, i.e.
\begin{eqnarray}
  X(p) = 
  \left({p^2 \over \Lambda^2}\right)^{\lambda \over 4\pi} ,
  \quad
  Y(p) = M X(p),
  \quad
  S(p) = {\hat{p}+M \over p^2} X(p) .
  \label{thsol}
\end{eqnarray}
\par
When there is no dynamical mass generation for the
fermion
$Y(p) = 0$  (i.e. $M = 0$), the wavefunction
renormalization function $A(p)$ has a solution given by
\begin{eqnarray}
 A_0(p) = X^{-1}(p) =
 \left({p^2 \over \Lambda^2}\right)^{-{\lambda \over
4\pi}} 
= \exp \left[-{\lambda \over 4\pi} 
\ln {p^2 \over \Lambda^2} \right].
\end{eqnarray}
Therefore, the SD equation (\ref{SD3}) has a consistent
solution (\ref{thsol}) which corresponds to no dynamical
mass generation $M=0$.
\par
Next, we want to ask whether the SD equation has a solution
of the dynamical mass generation, i.e. 
$M\not=0$ even for $m_0=0$.
In two dimensions, there is no spontaneous breakdown of
{\it continuous} symmetry due to
Mermin-Wagner \cite{MW61} and Coleman's theorem
\cite{Coleman73}.  Therefore, the U(1) chiral symmetry
$\Psi \rightarrow e^{i\gamma_5 \alpha} \Psi$ is not broken
spontaneously and the chiral order parameter vanishes
identically in two dimensions. 
This does not necessarily imply that the fermion remains
massless.  Actually, it is possible
for the fermion to acquire a mass according to Witten
\cite{Witten78}. From the solution obtained above, we find
that the following identity (corresponding to
Eq.~(\ref{op})) holds
\begin{eqnarray}
  M = \lambda  \int_{0}^{\Lambda^2} {dp^2 \over 4\pi} 
 {M \over p^2} X(p) ,
 \quad
 X(p) = \exp \left[ {\lambda \over 4\pi} 
\ln {p^2 \over \Lambda^2} \right] .
\label{mfeq}
\end{eqnarray}
If we truncate this equation up to some finite order of
$1/N_f$ expansion where $N_f G$ is fixed to be a constant
(i.e. $\lambda=O(1/N_f)$), we can get the non-trivial
scaling law. Actually, in the leading order of
$1/N_f$, we get the scaling of the essential singularity
type (at the origin $\lambda=0$):
\begin{eqnarray}
   {\mu \over \Lambda} = e^{-{2\pi \over \lambda}},
\end{eqnarray}
where we have introduced the infrared cutoff $\mu$ in
Eq.~(\ref{mfeq}) as a lower bound of integration which
plays the role of the dynamically generated mass.
This scaling shows the asymptotic freedom in two dimensions
\cite{GN74}.
\par
For finite $e^2$, finally, we find
\begin{eqnarray}
 && i e^2  \int {d^2k \over (2\pi)^2}
  \left[ {D_T(k^2) - L_\xi(k^2) \over k^2} \right]
 e^{-i k \cdot x}  
 \nonumber\\
 &=& - i \lambda_0 
 [ \Delta_F(x;0) - \Delta_F(x; e^2 \lambda_0^{-1})]
 + i G [\Delta_F(x;0) - \Delta_F(x; \xi G^{-1})] .
  \label{}
\end{eqnarray}
This shows that the difference of this equation with that
in the limit 
$e=\infty$ is just
$
i \lambda_0 \Delta_F(x; e^2 \lambda_0^{-1}) .
  \label{difference}
$
Hence Eq.~(\ref{SD2}) reads
\begin{eqnarray}
  X(p) &=&  1 +  \int {d^2q \over (2\pi)^2} 
  {(p-q) \cdot q \over q^2}
  \left[ {\lambda  \over (p-q)^2}
  - {\lambda_0 \over (p-q)^2 + e^2 \lambda_0^{-1}}
\right] X(q) ,
  \nonumber\\
  Y(p) &=&   \int {d^2q \over (2\pi)^2}  
  {p \cdot (p-q) \over q^2}
   \left[ {\lambda  \over (p-q)^2}
   - {\lambda_0 \over (p-q)^2 + e^2 \lambda_0^{-1}}
\right] Y(q) .
\label{SD5}
\end{eqnarray}
However, it is rather difficult to find the exact solution
for this equation.
In a subsequent paper \cite{Kondo96d}, we will give detailed
study on dynamical mass generation in the case of finite
$e^2$. Alternative approach to the gauged Thirring model is
the bosonization technique by which the
the apparent inconsistency (dynamical fermion
mass generation and asymptotic freedom without
breaking the continuous chiral symmetry) is clearly
understood, which will be given elsewhere
\cite{Kondo96e}.
Further studies of the gauged Thirring model will enable
us to compare the results with the lattice version of this
model investigated recently by Jersak et al.
\cite{Jersak96d=2}.

\section*{Acknowledgments}
The author would like to thank Prof. Ian J.R. Aitchison for
kind hospitality in Oxford.
This work is supported in part by the Japan Society for the
Promotion of Science and the Grant-in-Aid for Scientific
Research from the Ministry of Education, Science and
Culture (No.07640377).


\end{document}